# Penta-twinned gold nanoparticles under pressure: a comprehensive study

Camino Martín-Sánchez[a,b*], Ana Sánchez-Iglesias[c], José Antonio Barreda-Argüeso[b], Jean-Paul Itié[d], Paul Chauvigne[d], Luis M. Liz-Marzán[e,f], Fernando Rodríguez[b]

[a] Faculté des Sciences, Département de Chimie Physique, Université de Genève, 30 Quai Ernest-Ansermet, CH-1211 Genève, Switzerland
[b] MALTA Consolider, DCITIMAC, Facultad de Ciencias, University of Cantabria, Av. Los Castros 48, Santander, 39005, Spain
[c] Centro de Física de Materiales (CFM-MPC), CSIC-UPV/EHU, Paseo Manuel de Lardizabal 5, 20018 Donostia-San Sebastián, 20118, Spain
[d] Synchrotron SOLEIL, L'Orme des Merisiers St.Aubin, BP48, 91192 Gif-sur-Yvette, France
[e] CIC biomaGUNE, Basque Research and Technology Alliance (BRTA), Paseo de Miramón 194, Donostia-San Sebastián, 20014, Spain
[f] Ikerbasque, Basque Foundation for Science, Bilbao, 43018, Spain

Abstract

We report on the high-pressure optical and mechanical properties of penta-twinned gold nanoparticles (PT-AuNPs) of different geometries: decahedra, rods and bipyramids. Our results show that, unlike single-crystal (SC-AuNPs), PT-AuNPs preserve both their non-cubic crystal structures and their overall morphology up to 30 GPa. This structural integrity under compression is related to an enhanced mechanical resilience of PT-AuNPs, despite exhibiting bulk moduli comparable to those of SC-AuNPs. Notwithstanding, comparable pressure-induced localized surface plasmon resonance redshifts – for all nanoparticle geometries – were observed. Our analysis indicates that these shifts are primarily caused by changes in the refractive index of the surrounding medium, with electron density compression playing a minor role, contrasting with the behavior in SC-AuNPs, where electron density compression has a greater influence.

## 1. Introduction

Understanding the mechanical properties of metallic nanoparticles (NPs) is a fundamental challenge in nanoscience and nanotechnology, as these properties dictate the stability of NPs and their resistance to deformation under external stress. Moreover, mechanical deformation significantly affects the plasmonic response of NPs because their localized surface plasmon resonance (LSPR) and extinction coefficients are highly sensitive to changes in NP size and shape induced by external forces [1-3].

Recent studies have extensively explored the high-pressure behavior of single-crystal gold nanoparticles (SC-AuNPs), primarily rods and spheres, using optical extinction spectroscopy to track LSPR shifts, alongside with different X-ray and electron microscopy techniques to assess structural and morphological changes [1-14]. It is now well established that the face-centered cubic (*fcc*) structure of single-crystal AuNPs remains stable under hydrostatic conditions, with an equation of state similar to bulk gold. Additionally, their LSPR undergoes a redshift with increasing pressure, primarily due to the increase in refractive index of the surrounding medium, which counteracts the slight blueshift resulting from electron density compression.

In contrast, the high-pressure behavior of penta-twinned gold nanoparticles (PT-AuNPs) remains poorly understood. A key motivation for studying PT-AuNPs under high-pressure conditions is their non-cubic crystalline structure. Recent studies have shown that, depending on their shape, PT-AuNPs adopt either tetragonal or orthorhombic structures due to elastic strain from surface pressure [15-17]. Unlike *fcc* SC-AuNPs, PT-AuNPs exhibit slight distortions toward body-centered tetragonal (*bct*) or orthorhombic (*bco*) arrangements, raising questions about potential shape and volume changes under hydrostatic pressure, in turn inducing changes in the plasmonic properties. Furthermore, PT-AuNPs exhibit enhanced mechanical properties, including a greater resistance to tension, bending, and indentation 18-22]. However, their structural stability, possible shape modifications under high pressure, and equation of state –particularly their bulk modulus– remain unexplored despite being critical for understanding their mechanical behavior. AuNPs have recently been employed as sensors to measure solvent refractive indices under varying hydrostatic and non-hydrostatic pressures, leveraging their LSPR sensitivity to environmental pressure changes [8,9,11]. In this way, certain PT-AuNP geometries that exhibit strong plasmonic enhancement, such as bipyramids [23], emerge as promising candidates for highly sensitive high-pressure sensors.

We present a comprehensive study of the optical and mechanical properties of PT-AuNPs under high pressure. Using X-ray diffraction (XRD), optical extinction spectroscopy, and transmission electron microscopy (TEM), we examine pressure-induced changes in structure, morphology, and plasmonic response. By comparing different PT-AuNP geometries – nanorods (PT-AuRod), bipyramids (PT-AuBip), and decahedra (PT-AuDec)– we elucidate the influence of crystal structure and shape on the stiffness, deformation mechanisms, and plasmonic properties in fivefold twinned nanoparticles.

## 2. Materials and methods

Penta-twinned AuNP with three different geometries - rods, decahedra, bipyramids – with sizes ranging from 10 to 60 nm were synthesized *via* previously reported seeded-growth methods [24-26], followed by surface functionalization with thiolated poly(ethylenglycol), which provides colloidal stability in mixtures of short-chain alcohols, even under the application of high pressure [27]. The gold molar concentrations were set to around 10 mM to achieve suitable XRD diagrams for structural analysis [12,17]. **Figure 1** shows representative TEM images of the investigated PT-AuNP, as well as their respective dimensions and standard deviations.

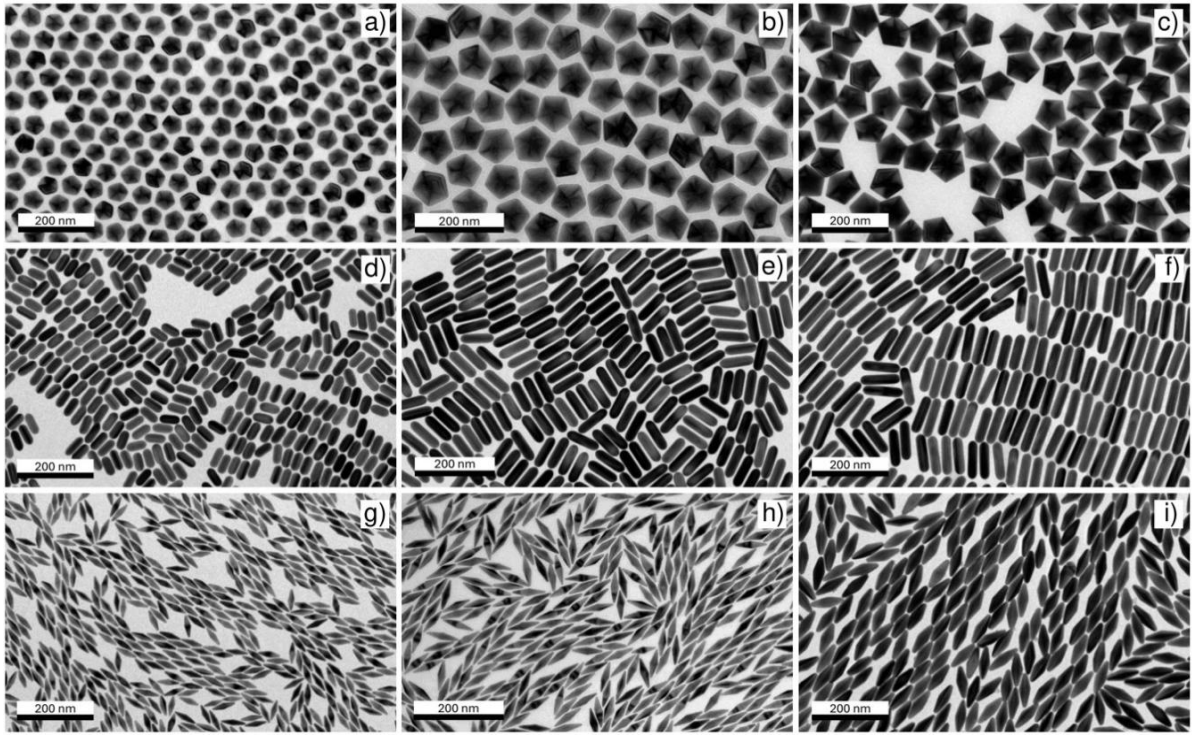

**Figure 1.** Representative TEM images of the PT-AuNP investigated in this work: a) (30 ± 1) nm PT-AuDec; b) (49 ± 1) nm PT-AuDec; c) (51 ± 1) nm PT-AuDec; d) (54 ± 2) × (25 ± 1) nm$^2$ PT-AuRod; (85 ± 3) × (26 ± 1) nm$^2$ PT-AuRod; f) (101 ± 3) × (26 ± 1) nm$^2$ PT-AuRod; g) (57 ± 1) × (17 ± 1) nm$^2$ PT-AuBip; h) (71 ± 3) × (19 ± 1) nm$^2$ PT-AuBip; i) (87 ± 3) × (30 ± 1) nm$^2$ PT-AuBip.

TEM images were obtained with a JEOL JEM-1400PLUS transmission electron microscope operating at an acceleration voltage of 120 kV. PT-AuNP colloids were measured before and after pressure treatments. In the latter case, the sample was recovered from the pressure chamber of the gasket by transferring the colloidal mixtures onto carbon-coated copper grids (400 square mesh) by touching the culet surface of the diamond anvil after pressure release.

XRD measurements on PT-AuNP MeOH-EtOH colloids were performed at the SOLEIL Synchrotron (France) using the PSICHÉ beamline. PT-AuNP colloidal solutions were loaded in a membrane diamond anvil cell (DAC) using a 35 μm-preindented rhenium gasket drilled with a 150 μm diameter hole. Compacted polycrystalline gold powder of 2 μm grain average size was used as a pressure calibrant. We followed the EOS reported by Heinz et al. [28], since the experimental measurements were carried out using MeOH-EtOH 4:1 as PTM which is similar to our system. This experimental methodology enables us to perform XRD measurements on both bulk and nanoscale gold under the same experimental conditions, which is crucial to get a precise comparison of lattice parameters in both systems. A parallel configuration geometry for diffraction (incident X-ray beam parallel to the DAC load axis) was used. 2D XRD data were collected on a CdTe2M Dectris detector, using a monochromatic X-ray beam with a wavelength of 0.3738 Å, focused to a beam size of 12×14 μm² (FWHM). The 2D XRD patterns were treated with the Dioptas program [29], and the resulting XRD patterns were analyzed using the Match! Software [30]. LeBail fits were accomplished using Lorentzian line profiles using two body centered cells with a tetragonal *I4/mmm* space groups for PT-AuDec, and an orthorhombic *Immm* space group for PT-AuBP and PT-AuRod. Instrumental resolution parameters were determined using of a sample $CeO_2$ of high crystalline quality using the Cagliotti equation [31]: U = -0.009421, V = -0.008602 y W = 0.002219, to account for the instrumental broadening: $B_{inst}^2 = W + V \tan\theta + U \tan^2\theta$. Subsequently, U, V and W were kept constant throughout the anaylisis. The lattice parameters were determined with an accuracy better than 0.003 Å, and the FWHM of the Bragg peaks, $B_r$, were determined with a precision of 0.001° using the equation $B_{exp}^2 = B_{inst}^2 + B_r^2$.

Optical extinction spectra at ambient conditions were recorded using a Cary6000i spectrophotometer with fused silica cuvettes (0.1 mm light path). High-pressure optical extinction spectra were collected in a home-built fiber-optic, two reflective objective-based microscope [32]. PT-AuNP MeOH-EtOH colloids were loaded in a Boehler-Almax DAC equipped with high transmission diamond anvils of 350 μm diameter culets. The pressure was applied in steps of approximately 0.3 GPa in upstroke and downstroke. To properly analyze the collected spectroscopic data, the baseline was corrected first, applying the same criterion for all sets of spectra, ensuring that the shape of the bands was unmodified. Then, the absorption bands were fitted to Pseudo-Voigt profiles on the wavelength scale.

## 3. Results and discussion

*3.1 X-ray diffraction under high-pressure conditions and transmission electron microscopy*

Figure 2 shows the variation with pressure of the XRD patterns for all three PT-AuNP geometries: PT-AuRod, PT-AuBip, and PT-AuDec. The room-temperature XRD patterns correspond to a tetragonal *I4/mmm* phase in decahedra, and an orthorhombic *Immm* phase in rods and bipyramids.

Notably, the respective tetragonal and orthorhombic structures remain intact under both hydrostatic and non-hydrostatic pressure conditions up to 20 GPa, unveiling the high stability of the penta-twinned nanoparticles with pressure, irrespective of the nanoparticle geometry. TEM images of the PT-AuNP colloids for the three geometries collected after application of high pressure confirm such a high stability. The images show that the shape and size distribution of the nanoparticles is maintained after pressure treatment. The sole observed difference between pristine and pressurized PT-AuBip is the apical vertices appearing marginally rounded in the pressurized particles, but their dimensions remain identical within statistical error, before and after the pressure treatment. This finding further reinforces the exceptional stability of both the crystal structure and habit of PT-AuNP under hydrostatic and quasi-hydrostatic pressures. Remarkably, these nanoparticles exhibit even greater resistance to deformation than SC-AuNR. When subjected to pressures exceeding 20 GPa, a significant fraction of SC-AuNR undergo fracture [13], whereas PT-AuNP remain unaffected.

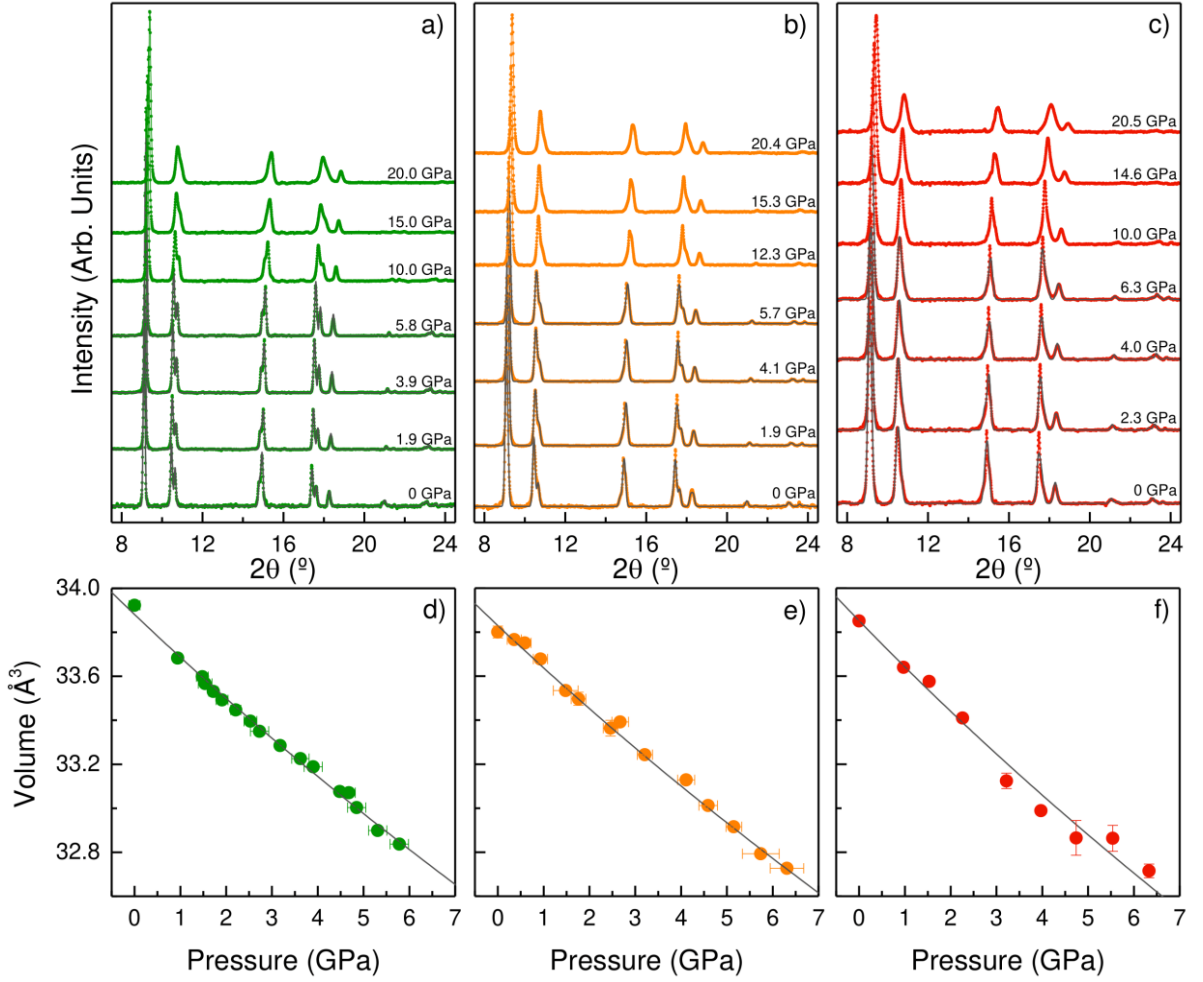

**Figure 2.** Selected diffraction patterns of a) PT-AuDec, b) PT-AuRod, and c) PT-AuBip, as a function of pressure. Dots correspond to experimental data; black lines correspond to a calculated XRD model in the hydrostatic regime. Pressure dependence of the cell volume for d) PT-AuDec, e) PT-AuRod, and f) PT-AuBip. Symbols correspond to experimental data; lines correspond to fits of the measured $V(P)$ data to the Vinet EOS. Error bars in both volume and pressure are either indicated or smaller than the symbols.

**Figure 2d-f** shows the variation of the lattice volume with increasing hydrostatic pressure. The $P(V)$ variation was described by a Vinet equation of state [33].

$$P = 3K_0 \frac{1-f}{f^2} \exp[\frac{3}{2}(K_0' - 1)(1 - f)] \qquad (1)$$

with $f = (V/V_0)^{1/3}$, where $K_0$ is the bulk modulus, $K_0'$ is the pressure derivative of the bulk modulus, and $V_0$ is the lattice volume at zero pressure. We used the same value of $K_0' = 5.72$ reported by Heinz et al. [28] in the fits for all three geometries to avoid parameter uncertainty and to allow comparison of bulk modulus and volume at zero pressure among all three samples.

The volume at zero pressure and the bulk moduli for the three investigated geometries are collected in **Table 1**. According to results shown in Table 1, the bulk moduli of PT-AuNP are similar to those of SC AuNP. The greatest discrepancy with respect to SC-AuNP is observed in PT-AuBip. It is worth noting that PT-AuBip are synthesized in the presence of $Ag^+$ to induce anisotropic growth. Inductively coupled plasma mass spectrometry (ICP-MS) measurements indicate that the content of silver in these PT-AuBip is ca. 3% [34]. Additionally, energy-dispersive X-ray spectroscopy (EDX) measurements indicate that silver atoms are prominently placed within the outer shell of the PT-AuBip [34]. Thus, although the main crystallographic structure of the nanoparticle is not significantly affected by the presence of silver atoms, the bulk modulus of the entire PT-AuBip might be reduced due to silver, as pure silver is characterized by a bulk modulus significantly smaller than that of gold (101 vs. 167 GPa, respectively) [35,36]. Additionally, the $V(P)$ data set corresponding to this geometry presents a lower number of experimental points, which leads to a significant uncertainty in the determination of the EOS. Overall, PT-AuNP exhibit the same bulk modulus as that of SC-AuNP within experimental accuracy. This result highlights that, regardless of their crystalline structure, both single crystal [12] and penta-twinned nanoparticles are slightly stiffer than bulk gold (Table 1).

**Table 1.** Fitting parameters using a Vinet equation of state [33] for the three different PT-AuNP geometries. The first derivative of the bulk modulus was fixed at $K_0' = 5.72$ [28]. Fit errors are given in brackets.

|  | $K_0$ (GPa) | $V_0$ (Å$^3$)/lattice cell | $V_0$ (Å$^3$)/Au |
|---|---|---|---|
| PT-AuDec | 171(4) | 33.881(13) | 16.9405(7) |
| PT-AuRod | 171(4) | 33.83(2) | 16.915(15) |
| PT-AuBip | 160(10) | 33.85(5) | 16.93(3) |
| Single crystal AuNP [12] | 170(3) | 67.643(18) | 16.911(4) |
| Single crystal bulk Au [12] | 167 (3) | 67.8524(6) | 16.9631(2) |

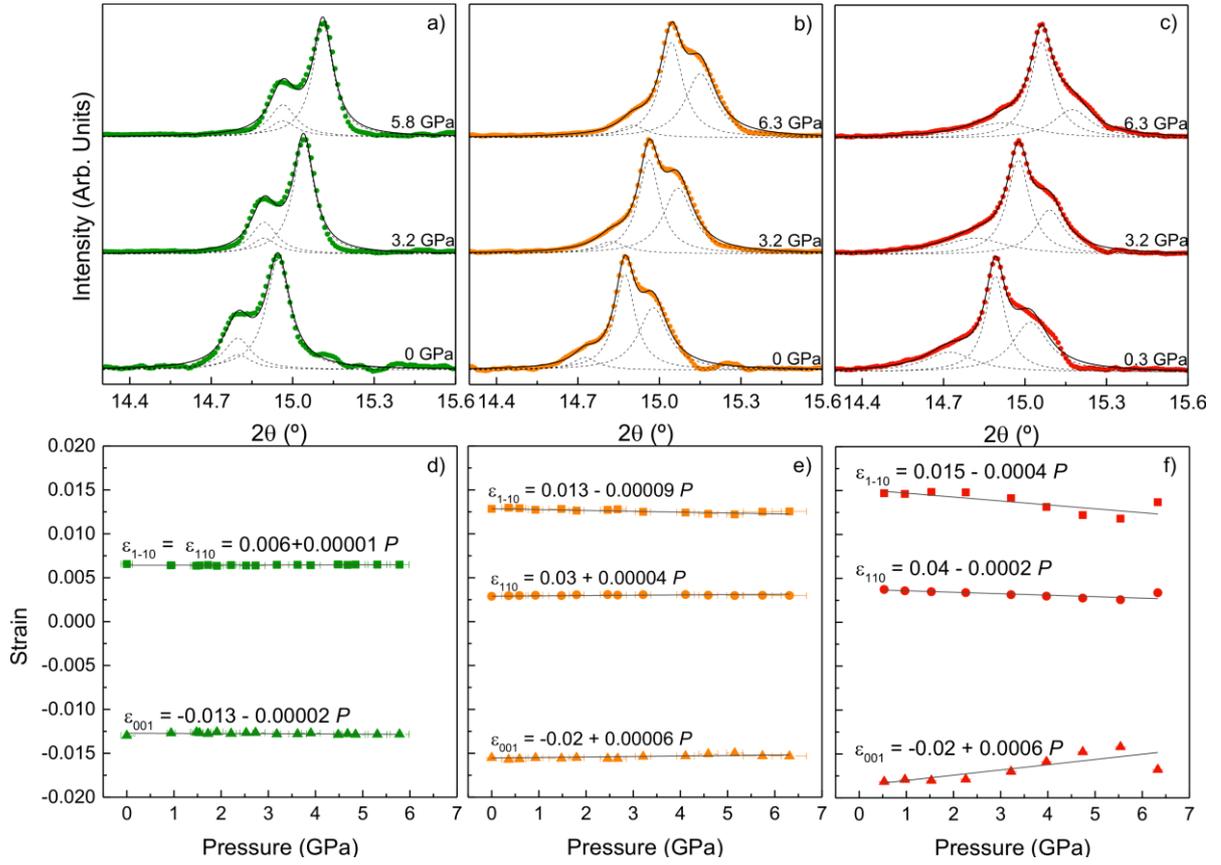

**Figure 3.** Cubic equivalent (220) Bragg peak for a) PT-AuDec, b) PT-AuRod, and c) PT-AuBip for selected pressures. Dots correspond to experimental data; solid lines correspond to Lorentzian profiles fitting to data; dashed lines indicate the deconvoluted peaks. Variation of the cubic lattice strains in d) PT-AuDec, e) PT-AuRod, and f) PT-AuBip with pressure. Strains are determined along the twin axis <110> and along the orthogonal directions <1-10> and <001> within the (110) plane. Filled symbols correspond to strain values derived from the cubic equivalent (220) Bragg peak; solid lines correspond to their least-square fits.

Shown in **Figure 3** is the variation of elastic strain along the three orthogonal directions $<110>_c$, $<1-10>_c$, and $<110>_c$ of the PT-AuNP, as a function of pressure (crystallographic directions refer to the cubic *fcc* phase). The strains of the PT-AuNP with respect to the cubic symmetry and their variation with pressure were obtained from the splitting of the *fcc* (220) Bragg peak, which is the most sensitive one to slight changes of the cubic symmetry in PT-AuNP, following the methodology described elsewhere [17]. It is worth noting that the peak splitting was maintained upon application of pressure, further supporting the excellent mechanic stability of PT-AuNP. Although initially the strain remained practically constant with pressure, as the hydrostatic pressure was further increased $\varepsilon_{110}$ and $\varepsilon_{001}$ slightly decreased and increased, respectively. These results indicate that particle shape essentially remains constant with pressure but PT-AuNP slightly changes in aspect ratio at high pressure. The relative aspect ratio, defined as δ*AR* / *AR*= $\varepsilon_{110} - \varepsilon_{001}$, varies as -0.08% GPa$^{-1}$ for PT-AuBip; -0.007% GPa$^{-1}$ for PT-AuRod; and +0.003% GPa$^{-1}$ for PT-AuDec in the hydrostatic pressure range up to around 6 GPa. In all cases, the aspect ratio can be considered constant within less than 1% accuracy in the whole hydrostatic pressure range. Such a variation cannot be detected by TEM and, as shown below, has no effect on the plasmonic behaviour under the application of pressure.

*3.2 Optical extinction measurements under high pressure conditions*

**Figures 4 and 5** show the optical extinction spectra for all three explored PT-AuNP geometries, with nanoparticles ranging in size from 10 to 60 nm, as a function of pressure. In all cases, the longitudinal LSPR (LLSPR) wavelength undergoes a redshift with increasing pressure. This redshift is significantly more pronounced in elongated nanoparticles (rods and bipyramids) compared to decahedra. While the redshift is approximately 100 nm for elongated geometries, it is around 60 nm for decahedra when pressure is increased from ambient to 20 GPa, in agreement with our previous report for SC-AuNP [3]. Upon pressure release, the spectra are nearly fully recovered, indicating an excellent reversibility and further highlighting the superior mechanical stability of PT-AuNP under pressure.

However, a distinct difference is observed between PT-AuNP and SC-AuNP in terms of the pressure dependence of the LSPR in nanorods. While LSPR(*P*) can be accurately described by the Gans theory for SC-AuRod within the hydrostatic pressure range, a slight deviation from this theory was observed for PT-AuRod, particularly for those with small aspect ratio. Figure 4 illustrates the variation of LLSPR of PT-AuRod as a function of pressure, along with predictions based on Gans theory [37]:

$$\lambda_{LSPR}(P) = \lambda_p \left(\frac{V_0}{V(P)}\right)^{1/2} \sqrt{\epsilon_{Au} + \frac{L}{L-1}\epsilon_m(P)} \qquad (2)$$

where $\lambda_p$ is the plasma frequency of gold at ambient pressure, $V_0$ and $V$ are the unit cell volumes of the nanoparticle at zero pressure and *P*, respectively, $\epsilon_{Au}$ is the high-frequency dielectric constant of gold at *P*, $\epsilon_m = n^2$ is the dielectric function of the surrounding medium at *P*, and *L* is the geometrical factor of the nanoparticle as defined elsewhere [38]. The parameters $\lambda_p$ and $\epsilon_{Au}$ were derived from the matching of the incompressible particle Gans model, i.e., $\frac{V_0}{V(P)} = 1$, to the empirical refractive index dependence of the LLSPR wavelength for each PT-AuRod aspect ratio [26]. The main difference between SC-AuRod and PT-AuRod lies in the deviation from the Gans model observed in the penta-twinned structure. The variation of the LLSPR wavelength in monocrystalline nanorods is accurately described by Eq. (2); however, in the case of penta-twinned nanorods, Eq. (2) underestimates the variation of the LLSPR wavelength with pressure. This deviation between the model and the experimental results becomes more significant for lower aspect ratios. For PT-AuRod, Eq. (2) requires the introduction of linear increase in $\epsilon_{Au}$ with pressure to accurately describe the LLSPR($P$) data. This behaviour, which is likely related to the unique structure of penta-twinned nanoparticles, warrants further investigation to provide a more comprehensive understanding.

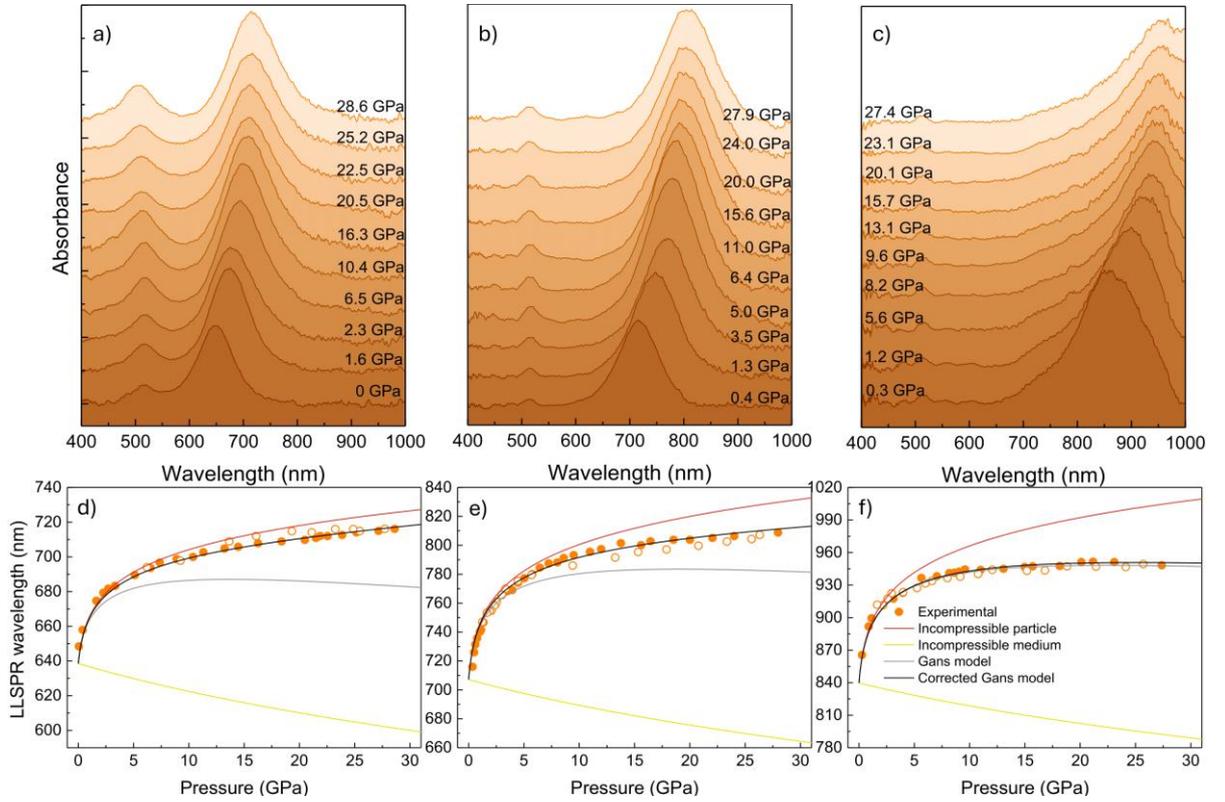

**Figure 4.** a-c) Raw extinction spectra of PT-AuRod with $AR = 2.2$ (a), $AR = 3.3$ (b), and $AR = 3.9$ (c), as a function of pressure. Spectra are vertically shifted for clarity. d-f) Pressure dependence of the LLSPR wavelength for PT-AuRod with $AR = 2.2$ (d), $AR = 3.3$ (e), and $AR = 3.9$ (f). Filled and empty circles correspond to experimental data taken in upstroke and downstroke, respectively. Lines represent the calculated LLSPR wavelengths for an incompressible medium (yellow line), incompressible particle (red line), compressible particle and medium (grey line), and compressible particle and medium and pressure-dependent $\epsilon_{Au}$ (black line).

The pressure dependence of the LLSPR of PT-AuBip and PT-AuDec is very similar to that of PT-AuRod. Although we have shown that plasmonic resonances of SC-AuNR and their pressure dependence can be well described like ellipsoids using Gans theory, in this work we have compared the experimental LLSPR pressure dependence for both PT-AuBip and PT-AuDec, with the LLSPR shift caused solely by the increase in the refractive index of the MeOH-EtOH 4:1 mixture under pressure. The latter contribution was determined based on the measured LLSPR shifts as a function of the surrounding refractive index of each geometry and size at ambient pressure. The results presented in **Figure 5** indicate that the LLSPR pressure variations for PT-AuBip and PT-AuDec are primarily influenced by changes in the refractive index of the surrounding medium, whereas the effects of increased gold electron density with pressure have a less significant impact.

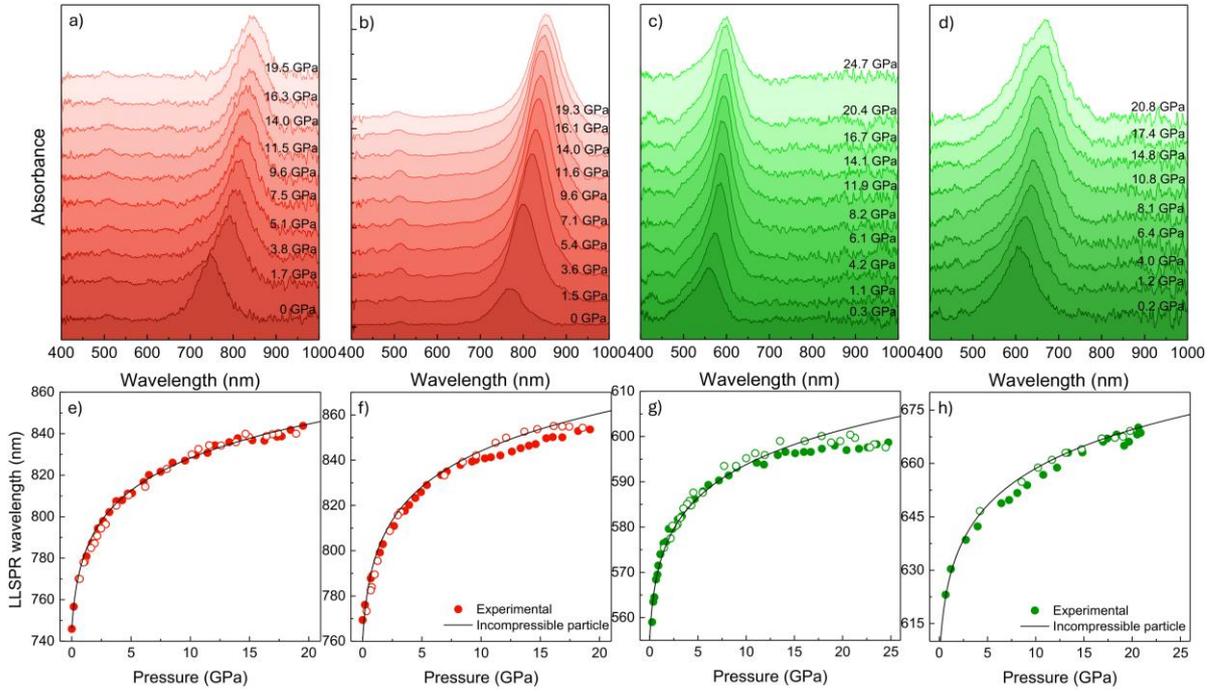

**Figure 5.** Raw extinction spectra of a) $AR = 2.9$ PT-AuBip, b) $AR = 3.3$ PT-AuBip, c) 30 nm PT-AuDec, and d) 50 nm PT-AuDec as a function of pressure. Spectra are vertically shifted for clarity. Pressure dependence of the LLSPR wavelength of e) $AR = 2.9$ PT-AuBip, f) $AR = 3.3$ PT-AuBip, g) 30 nm PT-AuDec, and h) 50 nm PT-AuDec. Filled and empty circles correspond to experimental data taken in upstroke and downstroke, respectively. Lines represent the calculated LLSPR wavelengths for an incompressible particle.

This apparently incompressible behaviour of PT-AuBip and PT-AuDec in their plasmonics response is similar to findings in PT-AuRod (see Figure 4). In this particular geometry, whose LLSPR pressure dependence can be described within the Gans theory, such an effect was related to an increase of the real part of the gold dielectric constant with pressure that compensated the decrease in the plasmon wavelength of gold by the increase in electronic density. Nevertheless, in the absence of an analytical model, this pressure-induced increase in the dielectric constant of gold in PT-AuDec and PT-AuBip requires additional confirmation.

## 4. Conclusions

Our results demonstrate that penta-twinned gold nanoparticles exhibit exceptional mechanical stability under pressure. Both tetragonal (decahedra) and orthorhombic (rods and bipyramids) structures, as well as the overall nanoparticle shape, remain remarkably stable up to a maximum applied pressure of 30 GPa. This stands in stark contrast to single-crystal nanospheres and nanorods, which undergo significant deformations or even fracture under similar conditions. These findings highlight the superior mechanical stability of penta-twinned nanoparticles, despite their bulk modulus being similar to single-crystal nanoparticles.

Regarding plasmonic properties, we observe a redshift in the localized surface plasmon resonance wavelength of penta-twinned nanoparticles. The magnitude of this redshift is comparable to that observed in single-crystal nanospheres and nanorods, with approximately 100 nm shifts for penta-twinned nanorods and bipyramids, and around 50 nm for decahedra. This suggests that elongated nanoparticles exhibit larger pressure-induced redshifts than spheroidal nanoparticles. Our analysis indicates that the shape and magnitude of these shifts are primarily attributed to changes in the refractive index of the surrounding medium, while the effects of electron density compression have a less significant impact, likely due to pressure enhancement of the real dielectric susceptibility of gold.

# Acknowledgments

Financial support from Projects PID2021-127656NB-I00, PID2023-151281OB-I00, and MALTA-Consolider Team (RED2018-102612-T) from the State Research Agency of Spain, Ministry of Science and Innovation is acknowledged. C.M.-S. acknowledges funding from the Spanish Ministry of Universities and the European Union-NextGeneration EU through the Margarita Salas research grant (C21.I4.P1). A.S.-I acknowledges the financial support received from the IKUR Strategy under the collaboration agreement between the Ikerbasque Foundation and Materials Physics Center on behalf of the Department of Science, Universities and Innovation of the Basque Government. We acknowledge SOLEIL for the provision of synchrotron radiation facilities, and we would like to thank the staff for assistance in using beamline PSICHÉ (proposal 20230273).

## Author Contributions

The manuscript was written through contributions of all authors. C.M.-S., L.M.L.-M and F.R. conceived the idea. A.S.-I synthesized and characterized the nanoparticles. C.M.-S., J.A.B.-A., J.-P.I., P.C. and F.R. conducted the X-ray diffraction experiments. C.M.-S. and F.R. performed the data analysis. C.M.-S. and F.R. wrote the original draft. All authors discussed the results and review the manuscript. C.M.-S. and F.R. coordinated and supervised the project.